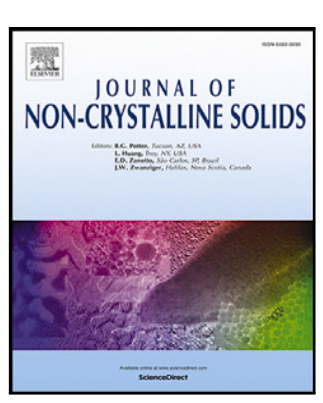

# New models of clean and hydrogenated amorphous silicon surfaces

K. Nepal [a],*, A. Gautam [a], C. Ugwumadu [b], D.A. Drabold [a]

[a] Department of Physics and Astronomy, Nanoscale and Quantum Phenomena Institute (NQPI), Ohio University, Athens, 45701, OH, USA
[b] Physics of Condensed Matter and Complex Systems (T-4) Group, Los Alamos National Laboratory, Los Alamos, 87545, NM, USA

ABSTRACT

We present new atomistic models of amorphous silicon (a-Si) and hydrogenated amorphous silicon (a-Si:H) surfaces. The a-Si model included 4096 atoms and was obtained using local orbital density functional theory. By analyzing a slab model (periodic in two dimensions with a slab about 44 Å thick), we observed a strong correlation between surface structure and surface charge density, which might be compared to STM experiments. Hydrogen atoms added near the under-coordinated surface atoms passivate dangling bonds and induce structural rearrangements. We analyze the electronic structure, including the localization of the states, and note resonant mixing between bulk and surface defect structures. We also compute the classical normal modes of the hydrogenated a-Si and compare them to experiments where possible. Our work is a step towards understanding the meaning of “surface reconstruction” for a noncrystalline material.

## 1. Introduction

Hydrogenated amorphous silicon (a-Si:H) is an important electronic material with applications ranging from night-vision devices to photovoltaics. It has been studied for decades and is in many ways, a mature material. One feature of a-Si:H that is incompletely understood is the surface of the material. In this paper, we present new computer models of a-Si and a-Si:H surfaces that elucidate interesting features about the surface topography (roughness), electronic activity (because of surface defect atoms), and the vibrations of the material in the harmonic approximation.

Extensive studies have been reported for bulk a-Si and a-Si:H [1–7]. In recent decades cSi/a-Si:H heterostructures have advanced the use of these materials as heterojunctions for efficient solar applications [8–10]. These studies characterize defects and explore the interface of silicon and hydrogen passivation to understand the electronic properties of silicon-based devices. Relatively little work focuses on a-Si and a-Si:H surfaces [11–15]. Hadjisavvas et al. [16] used Monte-Carlo simulations with empirical interatomic potentials to study large Si surface models. DFT simulations were implemented offering accurate descriptions of the a-Si:H surfaces, detailed in Ref. [11,17] for a small 216-atom slab model. The microscopic characterization of a-Si and a-Si:H surfaces can benefit from the large-scale DFT models but is computationally intensive. Recent advancements in DFT-based machine learning potentials have been implemented to bulk Si and Si:H [18–20], but the complexity at the surface demands DFT-level methodologies to accurately relax and characterize surface models.

We simulate surface models of clean amorphous silicon and hydrogenated amorphous silicon starting with an entirely four-coordinated 4096-atom bulk model of Djordjevic and Thorpe [21]. We obtain a clean a-Si surface by truncating periodic boundary conditions in one dimension, annealing, and relaxation. Next, a slab model for a-Si:H is obtained by adding hydrogen near the surface dangling bonds. The structural, electronic, and vibrational signatures of the as-formed a-Si and a-Si:H surface model are detailed.

The slab construction using this procedure may not fully replicate the true characteristics of the surface. The annealing process, performed at 1000 K for 2 ps, may not be sufficient to achieve complete surface relaxation, especially at newly created surfaces where under-coordination is present. A longer annealing time and/or different thermal conditions would ensure the surface reaches a true minimum-energy configuration. However, the models we obtained are stable, and the result was a slab of surface area $\approx$ 43.99 $\times$ 44.18 Å$^2$, a substantial advance over previous calculations (see Ref. [11]), providing new insights into surface behavior.

## 2. Computational details

The calculations were based on the local basis density functional code: the Spanish Initiative for Electronic Simulations with Thousands of Atoms (SIESTA) [22]. We employ a single-zeta basis set for Si and double-zeta for H and invoke the Harris functional to approximate the total energy and forces within the local density approximation (LDA).

* Corresponding author.
*E-mail addresses:* kn478619@ohio.edu (K. Nepal), drabold@ohio.edu (D.A. Drabold).

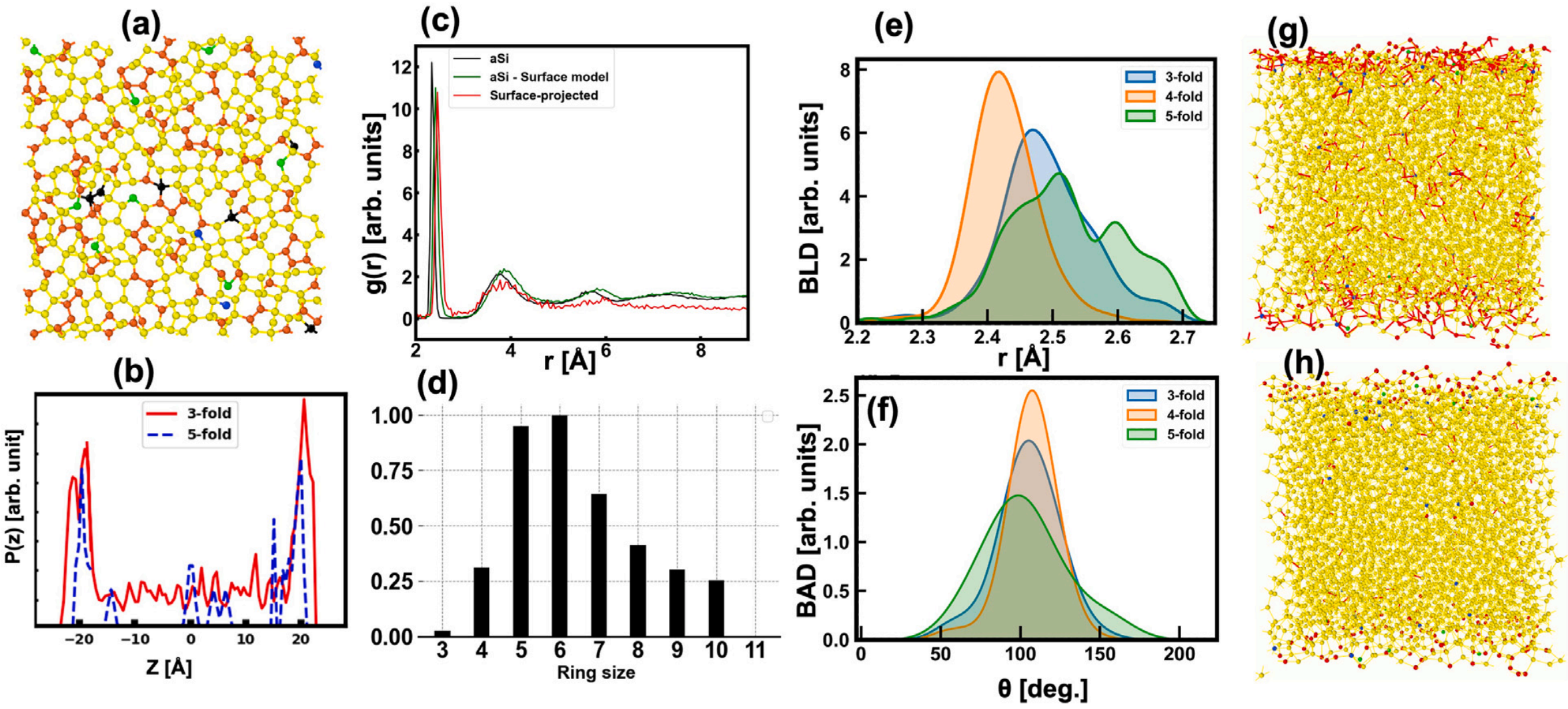

**Fig. 1.** (a) Top view for clean a-Si. Atoms within 5.3 Å from the topmost atom are shown. (b) Position probabilities of threefold-coordinated (red solid lines) and five-fold coordinated (blue dashed line) atoms as a function of the distance along the $z$-axis from the center of the cell (shifted to 0) towards the surface. (c) Radial distribution function ($g(r)$) (d) Ring distribution at the surface. (e–f) depicts the first neighbor bond length and angle distribution for different coordination in the a-Si surface slab model. (g–h) Side view for the a-Si surface slab model. Long bond-length ⟩ 2.5 Å and short bond-lengths ⟨ 2.32 Å, shown by red color. Most non-fourfold coordinated Si-atoms exhibit long but no short bond lengths. Blue, green, red, yellow, and black spheres show coordination 1–5, respectively. A 2.7 Å cutoff is used for bond coordination analysis, taken from the first minimum in the Si–Si radial distribution function.

**Table 1**
Coordination of silicon atoms in the amorphous silicon slab model and the surface only. A cut-off of 2.7 Å is used for coordination analysis taken from the first minimum in the radial distribution function.

| | Coordination | No. of Atoms | (%) |
|---|---|---|---|
| | 1 | 4 | 0.10 |
| | 2 | 21 | 0.51 |
| **a-Si-Slab** | 3 | 221 | 5.40 |
| | 4 | 3821 | 93.25 |
| | 5 | 29 | 0.71 |
| | 1 | 4 | 0.12 |
| | 2 | 21 | 2.55 |
| **Surface only** | 3 | 208 | 25.30 |
| (824) | 4 | 577 | 69.93 |
| | 5 | 14 | 1.70 |

**Table 2**
Pair correlation for Si–Si and Si–H bonds in various models comparing peaks (I, II, III). A cut-off of 2.7 Å is used for coordination analysis taken from the first minimum in the radial distribution function.

| g(r) | Total/surface-projected | | | Si-H | | |
|---|---|---|---|---|---|---|
| Peaks | I | II | III | I | II | III |
| a-Si | 2.35 | 3.75 | 5.74 | – | – | – |
| a-Si-Surf | 2.42/2.46 | 3.88/3.90 | 5.86/5.95 | – | - | – |
| a-Si:H-Surf | 2.42/2.44 | 3.87/3.88 | 5.88/5.95 | 1.65 | 3.3 | 5.25 |

A similar scheme has been applied efficiently to describe amorphous silicon and amorphous carbon [11,23,24].

A defect-free continuous random network cubic supercell containing 4096 atoms due to Thorpe [21] was relaxed with a conjugate gradient scheme as implemented in SIESTA. No significant rearrangements are observed for this model. To generate a surface model, periodicity was removed along the $z$-axis, creating a slab geometry with two surfaces. Of course, many atoms at the new surface are under-coordinated. The slab was annealed at 1000 K for 2 ps and relaxed anew using the conjugate gradient algorithm (with a force tolerance below 0.01 eV/Å), reaching a minimum energy configuration. This allows for an amorphous analog of surface reconstruction, yielding two stable surfaces. Properties reported here are averaged over the two surfaces, except where noted otherwise.

## 3. Clean amorphous silicon surface

We arbitrarily identify one surface as "top" and the other as "bottom". Both surfaces are taken as a slab with thickness 5.3 Å, taken from the top-most atom and the lower-most Si atom, respectively, consistent with the choice in Ref. [11]. This chosen surface depth encompasses the region where structural deviations from the bulk occur due to coordination defects. The number of surface atoms is 824 (391 at the top surface and 433 at the bottom surface). A wide range of coordination is displayed at the surface. Fig. 1(a) shows the configuration of surface silicon atoms, where different colored spheres correspond to different coordination: yellow fourfold, red threefold, green twofold, blue onefold, and black fivefold coordination.

The model interior is almost perfectly fourfold, and this coordination drops when approaching the surfaces (to ≈ 70%). Approximately 90% of the non-fourfold coordinated silicon atoms are within the top or bottom surface layers. 4 one-fold and 21 two-fold coordinated defective sites appeared on the surfaces. To estimate the distribution of non-tetrahedral coordination in the system, the positional probabilities ($P(z) = dN/dz$) *i.e.*, the number of non-fourfold coordinated atoms with distance along the $z$-axis is computed and shown in Fig. 1(b). The coordination statistics for the surface of a-Si are summarized in Table 1.

The total and surface-projected radial distribution function $g(r)$ was calculated and compared to that of a-Si. The calculated functions are shown in Figs. 1 (c) (total and surface projected are represented by green and red plots). While the total $g(r)$ shows a nearest-neighbor peak at ≈ 2.42 Å (which is a shift by ≈ 0.7 Å compared to a-Si, shown by black plots), surface-projected $g(r)$ exhibits greater shifts in radial distribution peaks. The first nearest neighbor peak in surface-projected $g(r)$ is obtained at 2.46 Å. Table 2 summarizes the positions of peaks in total pair correlation functions in our models. Next, the distribution of ring sizes at the reconstructed surface was analyzed. The silicon rings on the surface vary from 3 to 10 members, with the majority of silicon forming 5 to 7 members (Fig. 1(d)).

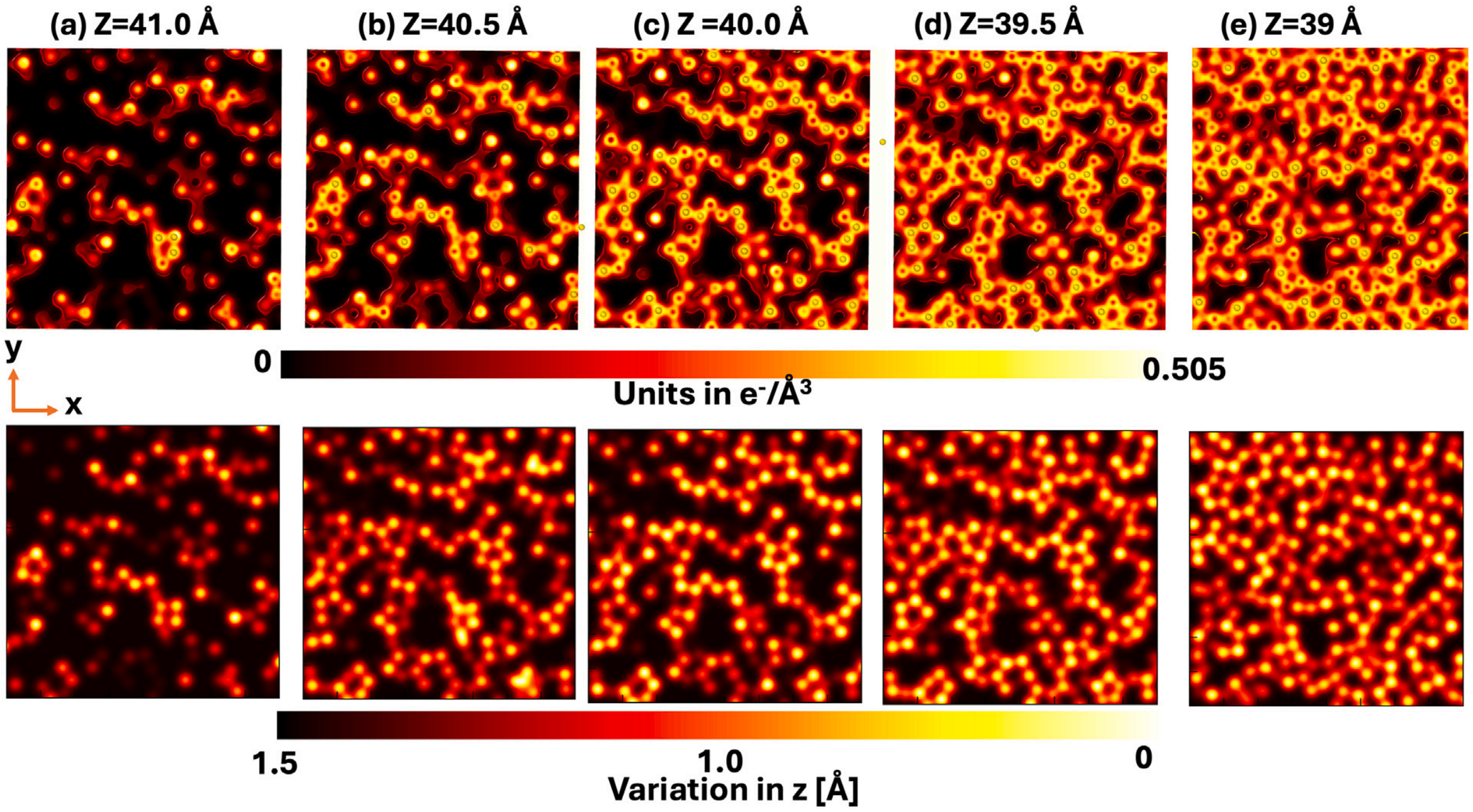

**Fig. 2.** [Upper row] Cross-sections of the charge density of clean silicon surface slices (thickness 0.5 Å) from the top showing the charge distribution progression inwards. [Lower row] Variation in the position of the atoms for the same planes as a heat map. Note the striking similarity of these physically distinct quantities.

To get deeper insights into these surface structures, we investigate the local bonding environment of silicon atoms at the surface. We computed bond length (BLD) and bond angle (BAD) distributions, identifying the variations in nearest neighbor Si–Si distances and Si–Si–Si bond angles for different coordination classes. Fig. 1 (e–f) displays BLD and BAD distributions for the a-Si surface model. Fourfold coordinated silicon atoms peak at approximately 2.42 Å and 106.7° with standard deviations of approximately 0.1 Å and 15.80 ° respectively (see Figs. 1(e–f) orange curves), demonstrating significant variation from the tetrahedral a-Si (2.35 Å bond length and 109.5° bond angle). Furthermore, the distribution of the bond length for the three-fold (blue curves) and five-fold (green curves) atoms (Fig. 1(e)) suggests that Si–Si forms larger bond lengths. Fig. 1(f) shows the bond angle analysis for 3- and 5-folded silicon atoms, shown in blue and green curves. These long bond lengths are concentrated at the surface and sparsely distributed in the bulk regions. Dangling and floating bonds form longer bond lengths (shown in red). In contrast, short bond lengths are associated with fourfold silicon atoms in bulk but are not associated with coordination defects. These observations are shown in Fig. 1 (g–h). The larger Si–Si bond length distribution is consistent with the shift in the first peak in surface projected $g(r)$.

Next, the surface charge density distribution was computed. We sliced the model into a thin shell of thickness 0.5 Å along the XY plane from the top surface. The charge distribution for each slice is shown in Fig. 2(b), depicting the local distribution of charges showing the progression from outer to inner slices. The color bar shows the intensity of charge density values increasing from dark to red hot. The plot showcases a network of silicon atoms with varying ring sizes at the surface. In the Figure, high-intensity (red-hot) and low-intensity (dark) regions indicate high and low charge densities at the cross-section of the plane.

To correlate the charge distribution with the spatial arrangement of silicon atoms (an indicator of surface roughness) at the surface, we calculated the variation in the z-coordinates of silicon atoms within a range of $\pm 1.5$ Å at different z-planes at the top surface. Fig. 2 [Lower row] shows the variation as a heat map, where the color bars represent atomic positions; atoms within $\pm$ 1 Å are colored in bright yellow to red, while those beyond this range appear in darker shades. A visual comparison for charge density and position variations, shown in Figs. 2 reveals a striking correlation between the surface charge density and the roughness of the silicon surface. The predicted charge densities should be comparable to an STM image (which we admit is probably not easy to obtain because of surface roughness and possibly issues of low electronic conductivity). The charge density and surface smoothness analysis performed in this study provides peak-to-valley variation in charge density at the surface, thus offering insights into the surface roughness of clean a-Si. We note that this study does not address the effect of annealing parameters (duration and higher temperature) on surface roughness, limiting insights into its correlation with thermal treatments.

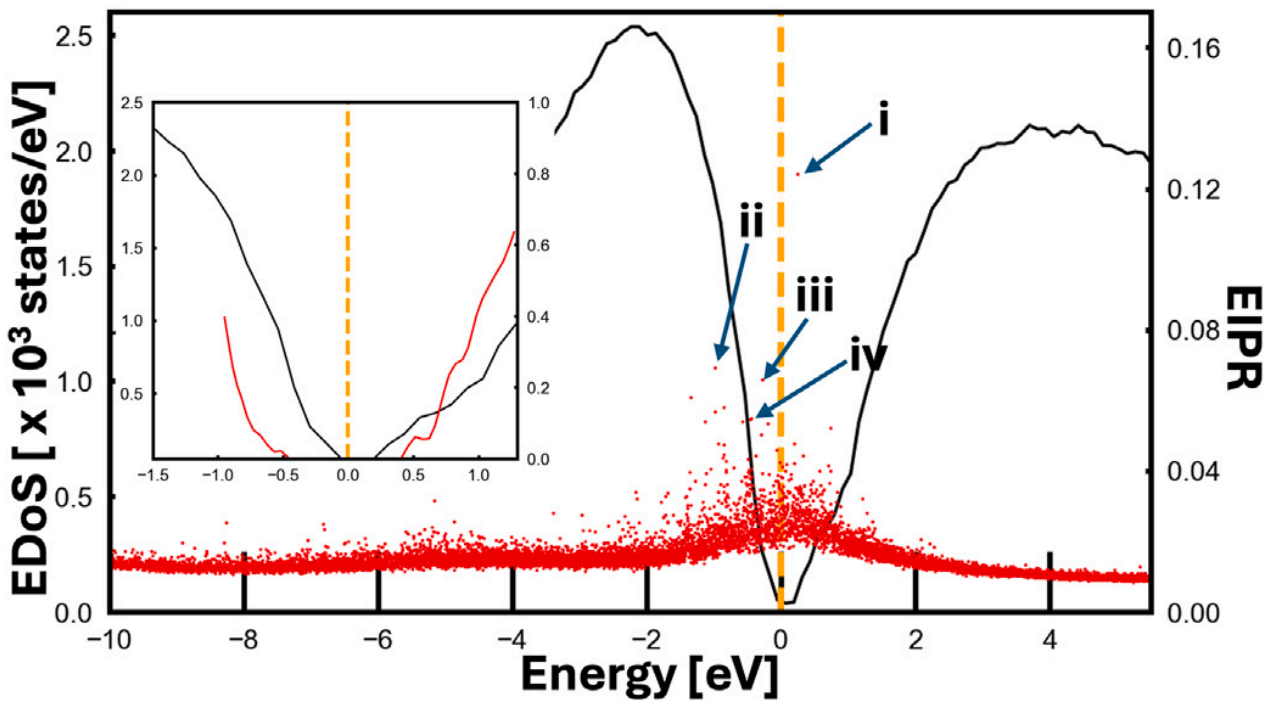

**Fig. 3.** Electronic DOS [black plots] and IPR [red dots] for clean amorphous silicon surface slab. The orange vertical dashed line is a Fermi level shifted to 0. The highlighted electronic states are discussed in the text. The valence and conduction tail states near the Fermi level (black plots) are shown in the inset compared to the bulk 4096 WWW amorphous silicon model [25] (red plot).

The electronic structure was investigated by analyzing the electronic density of states (EDoS) and inverse participation ratio (EIPR). The EDoS was calculated within SIESTA, using the Local Density Approximation (LDA) Harris functional with single-zeta basis set for Si and double-zeta basis set for H. An account of the localization of Kohn–Sham states ($\phi$) is given by EIPR defined as:

$$I(\phi_n) = \frac{\sum_i |c_n^i|^4}{(\sum_i |c_n^i|^2)^2} \quad (1)$$

where $c_n^i$ is the contribution to the eigenvector ($\phi_n$) from the $i$th atomic orbital as calculated with SIESTA. Low IPR values correspond to extended states (evenly distributed over atoms) and high IPR values to localized states.

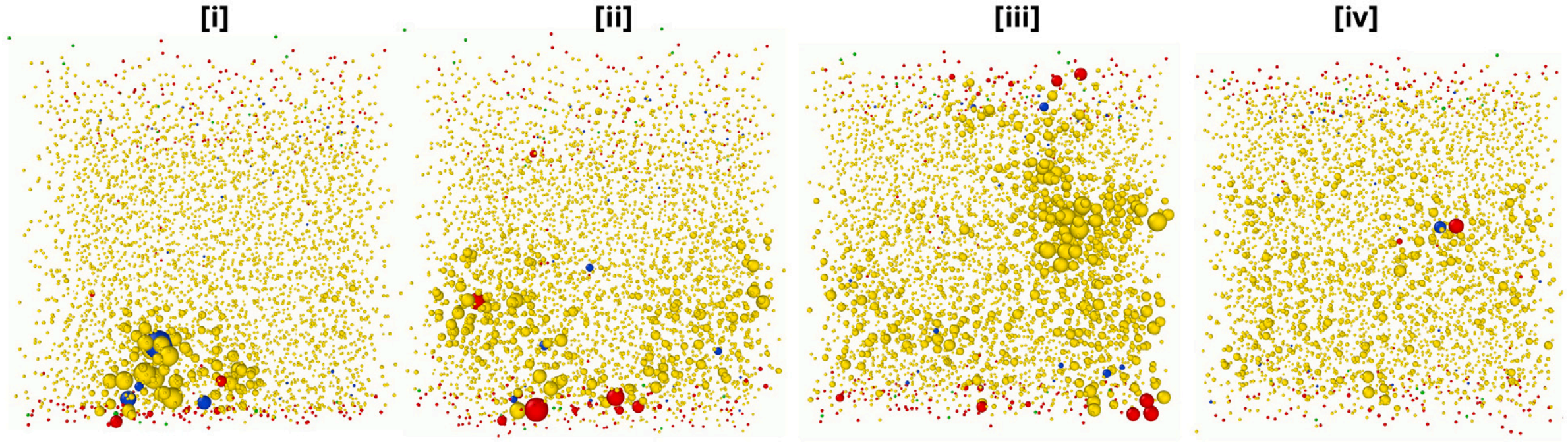

**Fig. 4.** Side view of clean a-Si surface slab model showing selected electronic states near the Fermi level projected onto the silicon atoms. The size of the spheres depicts the weight of the contribution of the silicon atom to given eigenstates. Green, red, yellow, and blue colors represent coordination in silicon atoms from 2–5 respectively.

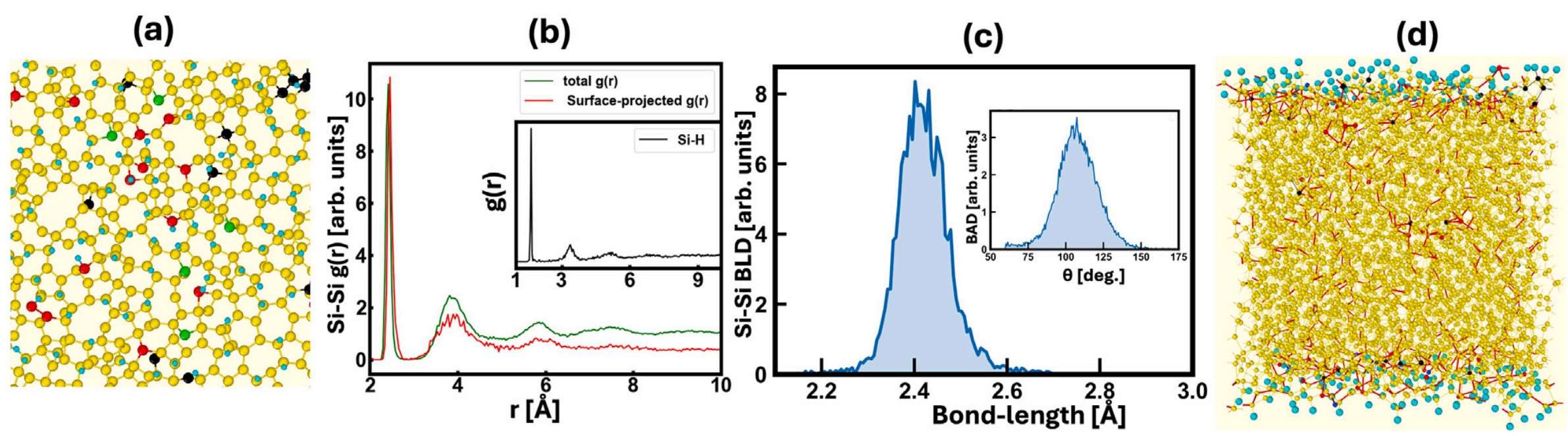

**Fig. 5.** (a) Top view for the hydrogenated a-Si surface. Atoms within 5.3 Å from the topmost atom are shown. A cutoff of 2.7 Å is used for bond identification. Blue, green, red, yellow, and black spheres identify the silicon coordination 2–5 sequentially, and cyan-colored spheres are hydrogen atoms. (b) Partial radial distribution functions. Figure (c) shows the bond length and bond angle distribution for silicon atoms. (d) Side view for the hydrogenated a-Si surface depicting bond length greater than 2.5 Å.

**Table 3**
Coordination of Si atoms in the hydrogenated a-Si surface. A cutoff of 2.7 Å is used for coordination analysis taken from the first minimum in the radial distribution function.

| | coordination | No. of Atoms | (%) |
|---|---|---|---|
| a-Si:H-Slab | 1 | 1 | 0.02 |
| | 2 | 11 | 0.27 |
| | 3 | 56 | 1.37 |
| | 4 | 3997 | 97.58 |
| | 5 | 31 | 0.76 |
| Surface only (840) | 1 | 1 | 0.12 |
| | 2 | 10 | 1.19 |
| | 3 | 37 | 4.40 |
| | 4 | 768 | 91.43 |
| | 5 | 24 | 2.86 |

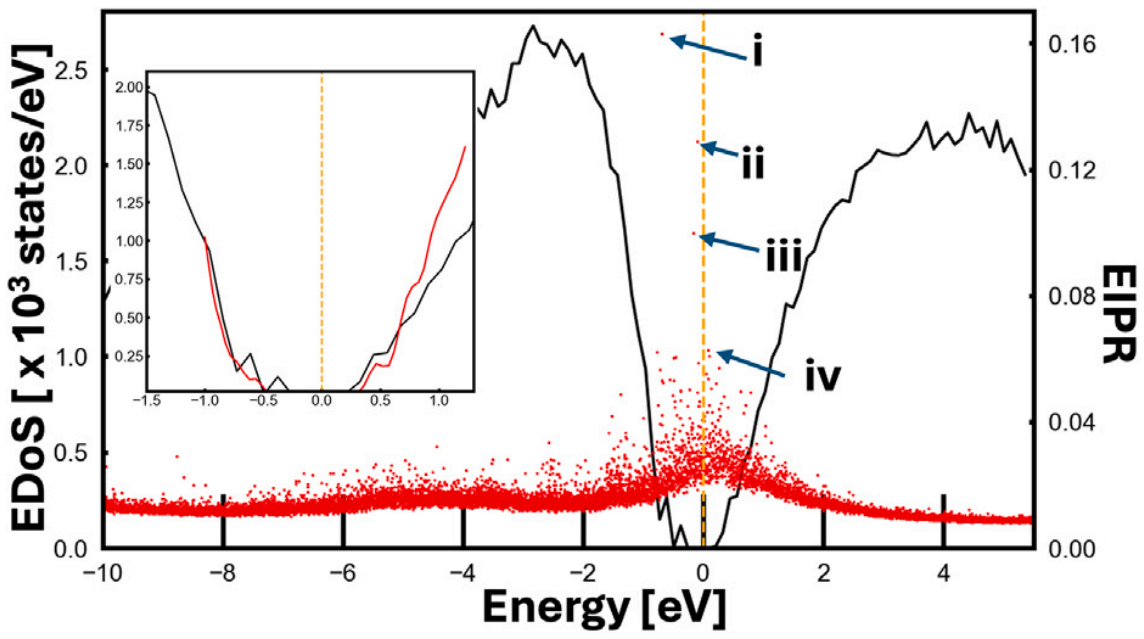

**Fig. 6.** Electronic DOS and IPR for hydrogenated a-Si surface. The vertical dashed line (orange) and conduction tail states near the Fermi level (black plots) are shown in the inset compared to the bulk 4096 WWW amorphous silicon model [25].

The EDOS for the clean a-Si surface is shown in Fig. 3. The EDoS spectrum shows a noticeable energy gap with states with some degree of localization, as shown by the higher inverse participation ratio (IPR) values in Fig. 3. These states near the Fermi level are attributed to surface defects. In the inset, the valence and conduction tails of the a-Si surface are compared to the bulk a-Si, using the 4096 atom amorphous silicon model by WWW, calculated by Igram et al. [21,25,26], shown by red plots. The bandtails are of course strongly affected by the surface defects. Previous studies [27,28] in bulk systems have demonstrated that these valence and conduction tail states are further sensitive to variations in bond lengths, both long and short associated with conduction and valence tails.

The relationship between the structural defects and the electronic density of states was analyzed by projecting several electronic states with high IPR values, labeled "i–iv" in Fig. 3, onto the silicon atoms. No strong localization was observed on surface dangling or floating bond atoms; however, the contribution to these states originates from defective surface regions and, in some cases, extends into the bulk. The electronic state labeled "i", shown in Fig. 4 [i], primarily involves a group of silicon atoms, with the majority of the contribution from five-fold coordinated silicon atoms, and some contributions from four-fold and three-fold coordinated atoms in its locality. For state "ii", a major contribution arises from a surface dangling bond, which extends into the bulk silicon atoms (see Fig. 4 [ii]). For state "iii", the majority of the contribution comes from a group of four-fold coordinated silicon atoms in the bulk, with an extension to the surface dangling atoms, shown in Fig. 4 [iii], showing an interesting mixing between bulk and surface states as in the resonant cluster proliferation model [29,30]. For state "iv", the primary contribution arises from dangling and floating bonds in the bulk, shown in Fig. 4 [iv].

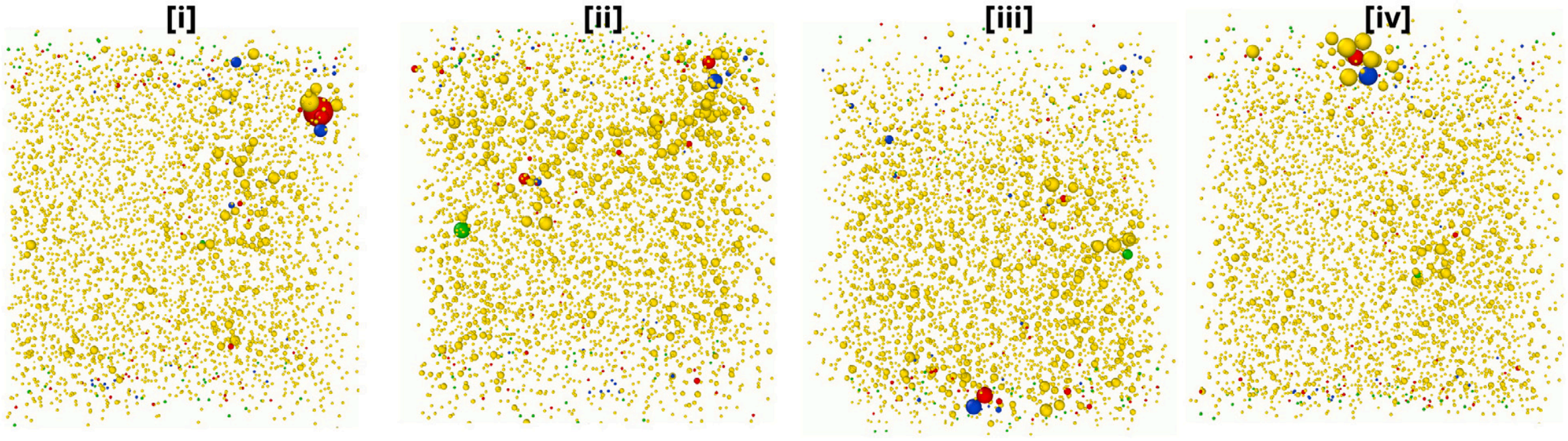

**Fig. 7.** Selected electronic states of a-Si:H near the Fermi level projected onto the silicon atoms. The size of the spheres depicts the weight of the contribution of the silicon atom to given eigenstates. Green, red, yellow, and blue colors represent coordination in silicon atoms from 2–5 sequentially.

## 4. Hydrogenated amorphous silicon surface

We added 207 hydrogen atoms positioned 1.5 Å above the dangling silicon atoms on both the top and bottom surfaces. The structure was relaxed to achieve an energy-converged structure. Fig. 5(a) shows the relaxed surface with hydrogen passivation, highlighting the hydrogen-induced structural changes. There indicates a dramatic reduction in defects at the surface (one-fold silicon atoms dropped from 4 to 1 and two-fold silicon from 21 to 11). The fraction of total fourfold coordinated atoms in the model increased to ≈ 98%. While the fraction of threefold-coordinated sites was reduced to 4.40% at the surface, four-fold coordination at the surface increased from ≈ 69.80% to ≈ 91.43%. The summary of the coordination statistics for the hydrogenated silicon surface is illustrated in Table 3.

The total and surface projected radial distributions for the hydrogenated amorphous silicon model are more consistent, as illustrated in Fig. 5(b). The first peaks in the radial distribution are observed at ≈ 2.42 Å and ≈ 2.44 Å, respectively. The partial $g(r)$ for silicon and hydrogen is shown in the inset, depicting the silicon hydrogen bond lengths peak at ≈ 1.65 Å. The position of peaks in a radial distribution for Si–Si and Si-H is tabulated in the last row in Table 2. With increasing four-fold coordination, the bond length and bond angle distributions peak at ≈ 2.40 Å and 107.65 ° with standard deviations of approximately 0.1 Å and 13.7 ° respectively, shown in Fig. 5(c). As illustrated in Fig. 5(d), a significant reduction in Si–Si long bond length is observed (≈ by 20%).

The EDoS and EIPR for the hydrogenated amorphous Si surface model are shown in Figs. 6. The electronic density of states is qualitatively similar to clean a-Si; however, the opening at the band gap is due to the reduction of the surface states. At the energy range of ± 1 eV from the Fermi level, the reduction of 113 states was observed. The electronic structure becomes more like that of bulk amorphous silicon. The inset shows the valence and conduction tails for the hydrogenated a-Si surface compared to that of the bulk for the 4096 WWW a-Si model [25]. Extensive studies have been made for such exponential tail (valance and conduction) in bulk amorphous silicon [31].

The electronic localization, shown with red drop-lines in Fig. 6, highlights the states near the band gap. As for clean amorphous silicon, these gap states derive contributions from defective surface atoms and sometimes extend into the bulk. The electronic state labeled "i" is centered around a dangling bond, with minor contributions from neighboring silicon atoms, as shown in Fig. 7 [i]. State "ii" primarily originates from under-coordinated silicon near the surface, but it extends into the bulk silicon atoms, as illustrated in Fig. 7 [ii]. Figs. 7[iii–iv] show states "iii" and "iv", which have their primary contributions from dangling and floating bonds at the surface, as well as from atoms in their neighborhood.

## 5. Vibrational properties of hydrogenated a-Si

The vibrational signatures of the hydrogenated a-Si surface were examined by computing the vibrational density of states (VDoS) and vibrational inverse participation ratio (VIPR) within the harmonic approximation (HO) and compared to the bulk a-Si. The Dynamical Matrix (DM) is constructed using:

$$D_{ij}^{\alpha\beta} = \frac{1}{\sqrt{m_i m_j}} \frac{\partial^2 E}{\partial u_i^\alpha \partial u_j^\beta} \tag{2}$$

Here, $u_i^\alpha$ is the small displacement of $i$th atom along cartesian ($\alpha$) direction, $m_i$ is the mass of the $i$th atom. The eigenvalue problem for the classical normal modes at the center (**k=0**) of the phonon Brillouin zone is [33]:

$$\omega_m^2 \; \mu_i^\alpha = \sum_{\beta j} D_{ij}^{\alpha\beta} \mu_j^\beta \tag{3}$$

where, $\omega_m$ and $\mu_i^\alpha$ are the vibrational frequency of the mode $m$ and the polarization of the vibrational mode $m$ at atom $i$ along $\alpha$ direction. VDOS is a vibrational frequencies spectrum, and the VIPR corresponding to the vibrational modes $m$ are computed using:

$$\zeta_m = \frac{\sum_n |\mu_n^m|^4}{\left(\sum_n |\mu_n^m|^2\right)^2} \tag{4}$$

The VIPR varies between 1/N (extended vibrational mode) and 1 (localized vibrational mode), providing the localization behavior of the normal vibrational modes.

Calculating the VDOS for our large system in SIESTA is computationally demanding, requiring a massive force constant matrix and of order 6 × 4303 finite displacements. Instead, we used the machine learning Gaussian Approximation Potential (ML-GAP) for hydrogenated a-Si [19], implemented within the "Large-scale Atomic/Molecular Massively Parallel Simulator" (LAMMPS) software package [34]. The slab model from SIESTA was energetically relaxed, and the force constant matrix was computed. After relaxation in LAMMPS, the change in energy per atom was ≈ 0.04 eV. Each atom was displaced by 0.01 Å in the x±, y±, and z± directions to compute the DM defined in (2).

The VDOS and VIPR for hydrogenated a-Si are shown in Fig. 8. The major peaks in VDOS lie below 550 cm$^{-1}$. The low-frequency VDOS for a-Si:H is compared to the VDOS for bulk a-Si calculated experimentally [32], and by Igram et al. [25], shows significant consistency as shown in Fig. 8(a). Smaller peaks throughout the frequency range of 600–1400 cm$^{-1}$ shown in Fig. 8 (b) are observed, attributed to different Si–H vibrational modes (wagging and bending modes) [35,36]. These higher frequencies vibrational modes are localized, depicted by higher IPR values. A distinct broader peak at the frequency 2060 cm$^{-1}$ is observed, shown in Fig. 8(c), attributed to the stretching modes of Si–H. These stretching modes are highly localized, consistent with Refs. [17,37], indicated by high VIPR values in Fig. 8(c).

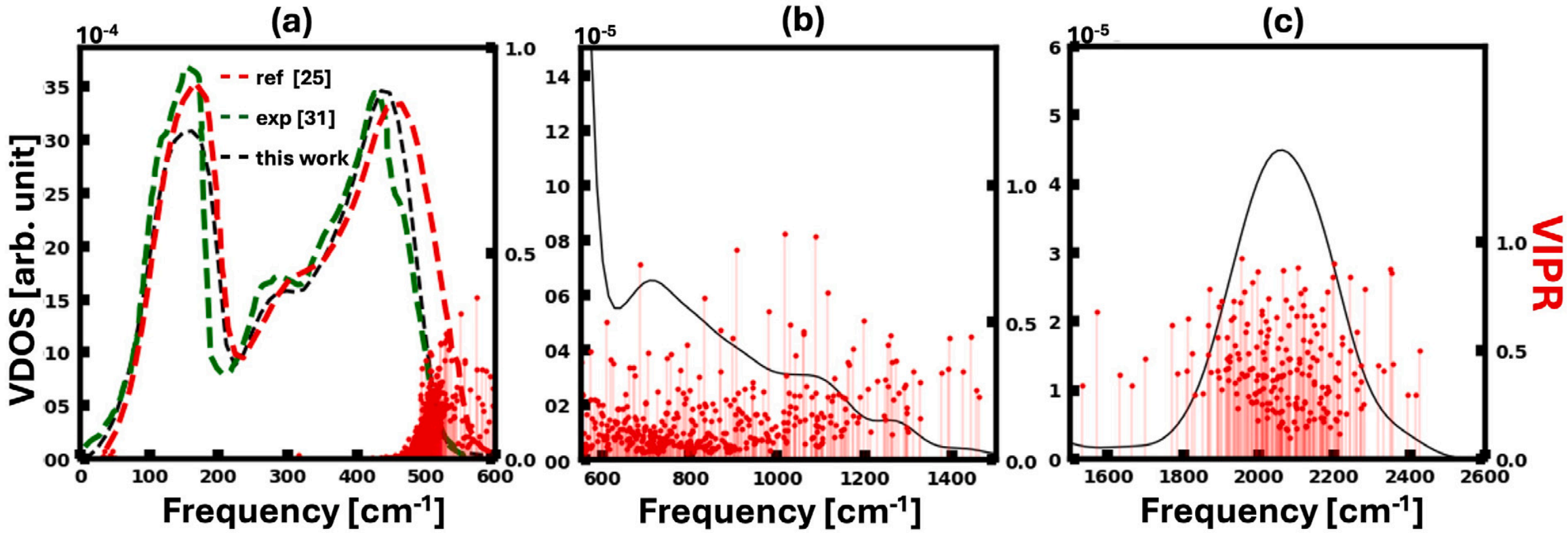

**Fig. 8.** Vibrational density of states and vibrational inverse participation ratio for hydrogenated amorphous silicon (a-Si:H) surface. Figures (a–c) show the VDOS and VIPR spectra over three energy ranges 0–600 cm$^{-1}$, 550–1500 cm$^{-1}$, and 1500–2600 cm$^{-1}$, respectively. The low frequencies VDOS (below 600 cm$^{-1}$) spectrum for a-Si:H is compared to the bulk a-Si (experiment [32] - green plots and theoretical study [25] - red plots).

## 6. Conclusion

This study presents atomistic models of a clean and hydrogenated amorphous silicon model of a fairly large surface area. A significant correlation between surface roughness and charge density for the clean a-Si surface model is noted, providing insights that can be compared to STM experiments. The electronic density of states at the Fermi level revealed electronic states localized at the surface defects and resonant mixing between surface and bulk structures.

Next, adding hydrogen to surface dangling bonds creates a band gap, modifying the material's stability and electronic properties. Vibrational analysis on a-Si:H further revealed that lower-frequency modes for a-Si:H are consistent with bulk amorphous silicon, while higher-frequency modes are localized due to Si-H interactions. This work provides models that will be useful for further investigation into these surfaces.

## CRediT authorship contribution statement

**K. Nepal:** Writing – original draft, Visualization, Validation, Software, Resources, Methodology, Investigation, Formal analysis, Data curation, Conceptualization. **A. Gautam:** Writing – review & editing, Visualization, Software. **C. Ugwumadu:** Writing – review & editing, Validation, Formal analysis, Conceptualization. **D.A. Drabold:** Writing – review & editing, Validation, Supervision, Software, Project administration, Funding acquisition, Conceptualization.

## Declaration of competing interest

The authors declare no competing interest.

## Acknowledgments

We acknowledge computational support from the U.S. National Science Foundation grant MRI 2320493.

## References

[1] C.G. Van de Walle, R. Street, Structure, energetics, and dissociation of Si-H bonds at dangling bonds in silicon, Phys. Rev. B 49 (20) (1994) 14766, http://dx.doi.org/10.1103/PhysRevB.49.14766.

[2] A. Valladares, F. Alvarez, Z. Liu, J. Sticht, J. Harris, Ab initio studies of the atomic and electronic structure of pure and hydrogenated a-Si, Eur. Phys. J. B-Condens. Matter Complex Syst. 22 (2001) 443–453, http://dx.doi.org/10.1007/s100510170094.

[3] A. Descoeudres, L. Barraud, S. De Wolf, B. Strahm, D. Lachenal, C. Guérin, Z. Holman, F. Zicarelli, B. Demaurex, J. Seif, et al., Improved amorphous/crystalline silicon interface passivation by hydrogen plasma treatment, Appl. Phys. Lett. 99 (12) (2011) http://dx.doi.org/10.1063/1.3641899.

[4] D. Drabold, T. Abtew, F. Inam, Y. Pan, Network structure and dynamics of hydrogenated amorphous silicon, J. Non-Cryst. Solids 354 (19–25) (2008) 2149–2154, http://dx.doi.org/10.1016/j.jnoncrysol.2007.09.081.

[5] M. Stuckelberger, R. Biron, N. Wyrsch, F.-J. Haug, C. Ballif, Progress in solar cells from hydrogenated amorphous silicon, Renew. Sustain. Energy Rev. 76 (2017) 1497–1523, http://dx.doi.org/10.1016/j.rser.2016.11.190.

[6] K. Jarolimek, R. De Groot, G. De Wijs, M. Zeman, First-principles study of hydrogenated amorphous silicon, Phys. Rev. B— Condens. Matter Mater. Phys. 79 (15) (2009) 155206, http://dx.doi.org/10.1103/PhysRevB.79.155206.

[7] J.D. Joannopoulos, G. Lucovsky, The physics of hydrogenated amorphous silicon II: Electronic and Vibrational Properties, 1, vol. 56, Springer Berlin, Heidelberg, 2008, http://dx.doi.org/10.1007/3540128077.

[8] M. Tosolini, L. Colombo, M. Peressi, Atomic-scale model of c-Si/a-Si:H interfaces, Phys. Rev. B 69 (7) (2004) http://dx.doi.org/10.1103/physrevb.69.075301.

[9] R.V. Meidanshahi, D. Vasileska, S.M. Goodnick, Role of Hydrogen in the Electronic Properties of a-Si:H/c-Si Heterostructures, J. Phys. Chem. C 125 (23) (2021) 13050–13058, http://dx.doi.org/10.1021/acs.jpcc.1c03173.

[10] M. Burrows, U. Das, R. Opila, S. De Wolf, R. Birkmire, Role of hydrogen bonding environment in a-Si: H films for c-Si surface passivation, J. Vac. Sci. Technol. A 26 (4) (2008) 683–687, http://dx.doi.org/10.1116/1.2897929.

[11] K.A. Kilian, D.A. Drabold, J.B. Adams, First-principles simulations of a-Si and a-Si:H surfaces, Phys. Rev. B 48 (1993) 17393–17399, http://dx.doi.org/10.1103/PhysRevB.48.17393.

[12] M. Nolan, M. Legesse, G. Fagas, Surface orientation effects in crystalline–amorphous silicon interfaces, Phys. Chem. Chem. Phys. 14 (43) (2012) 15173–15179, http://dx.doi.org/10.1039/C2CP42679J.

[13] J.R. Abelson, Plasma deposition of hydrogenated amorphous silicon: Studies of the growth surface, Appl. Phys. A 56 (1993) 493–512, http://dx.doi.org/10.1007/BF00331400.

[14] A. Dalton, E. Seebauer, Structure and mobility on amorphous silicon surfaces, Surf. Sci. 550 (1–3) (2004) 140–148, http://dx.doi.org/10.1016/j.susc.2003.12.033.

[15] C.S. Ewing, S. Bhavsar, G. Veser, J.J. McCarthy, J.K. Johnson, Accurate amorphous silica surface models from first-principles thermodynamics of surface dehydroxylation, Langmuir 30 (18) (2014) 5133–5141, http://dx.doi.org/10.1021/la500422p.

[16] G. Hadjisavvas, G. Kopidakis, P.C. Kelires, Structural models of amorphous silicon surfaces, Phys. Rev. B 64 (2001) 125413, http://dx.doi.org/10.1103/PhysRevB.64.125413.

[17] R. Singh, S. Prakash, N.N. Shukla, R. Prasad, Sample dependence of the structural, vibrational, and electronic properties of aSi:H: A density-functional-based tight-binding study, Phys. Rev. B 70 (11) (2004) http://dx.doi.org/10.1103/physrevb.70.115213.

[18] G. Csanyi, Research data: Machine learning a general-purpose interatomic potential for silicon, 2021, http://dx.doi.org/10.17863/CAM.65004.

[19] D. Unruh, R.V. Meidanshahi, S.M. Goodnick, G. Csányi, G.T. Zimányi, Gaussian approximation potential for amorphous Si : H, Phys. Rev. Mater. 6 (6) (2022) http://dx.doi.org/10.1103/physrevmaterials.6.065603.

[20] V.L. Deringer, N. Bernstein, A.P. Bartók, M.J. Cliffe, R.N. Kerber, L.E. Marbella, C.P. Grey, S.R. Elliott, G. Csányi, Realistic atomistic structure of amorphous silicon from machine-learning-driven molecular dynamics, J. Phys. Chem. Lett. 9 (11) (2018) 2879–2885, http://dx.doi.org/10.1021/acs.jpclett.8b00902.

[21] B.R. Djordjević, M.F. Thorpe, F. Wooten, Computer model of tetrahedral amorphous diamond, Phys. Rev. B 52 (8) (1995) 5685–5689, http://dx.doi.org/10.1103/physrevb.52.5685.

[22] J.M. Soler, E. Artacho, J.D. Gale, A. García, J. Junquera, P. Ordejón, D. Sánchez-Portal, The SIESTA method for *ab initio* materials simulation, J. Phys.: Condens. Matter. 14 (11) (2002) 2745–2779, http://dx.doi.org/10.1088/0953-8984/14/11/302.

[23] J. Dong, D.A. Drabold, Ring formation and the structural and electronic properties of tetrahedral amorphous carbon surfaces, Phys. Rev. B 57 (1998) 15591–15598, http://dx.doi.org/10.1103/PhysRevB.57.15591.

[24] S.H. Yang, D.A. Drabold, J.B. Adams, Ab initio study of diamond C(100) surfaces, Phys. Rev. B 48 (1993) 5261–5264, http://dx.doi.org/10.1103/PhysRevB.48.5261.

[25] D. Igram, B. Bhattarai, P. Biswas, D. Drabold, Large and realistic models of amorphous silicon, J. Non-Cryst. Solids (ISSN: 0022-3093) 492 (2018) 27–32, http://dx.doi.org/10.1016/j.jnoncrysol.2018.04.011.

[26] F. Wooten, K. Winer, D. Weaire, Computer generation of structural models of amorphous Si and Ge, Phys. Rev. Lett. 54 (13) (1985) 1392–1395, http://dx.doi.org/10.1103/physrevlett.54.1392.

[27] P. Fedders, D. Drabold, S. Nakhmanson, Theoretical study on the nature of band-tail states in amorphous Si, Phys. Rev. B - Condens. Matter Mater. Phys. 58 (23) (1998) 15624–15631, http://dx.doi.org/10.1103/PhysRevB.58.15624.

[28] Y. Pan, M. Zhang, D. Drabold, Topological and topological-electronic correlations in amorphous silicon, J. Non-Cryst. Solids 354 (29) (2008) 3480–3485, http://dx.doi.org/10.1016/j.jnoncrysol.2008.02.021.

[29] J. Dong, D.A. Drabold, Atomistic structure of band-tail states in amorphous silicon, Phys. Rev. Lett. 80 (1998) 1928–1931, http://dx.doi.org/10.1103/PhysRevLett.80.1928.

[30] J.J. Ludlam, S.N. Taraskin, S.R. Elliott, D.A. Drabold, Universal features of localized eigenstates in disordered systems, J. Phys.: Condens. Matter. 17 (30) (2005) L321–L327, http://dx.doi.org/10.1088/0953-8984/17/30/l01.

[31] D.A. Drabold, Y. Li, B. Cai, M. Zhang, Urbach tails of amorphous silicon, Phys. Rev. B 83 (2011) 045201, http://dx.doi.org/10.1103/PhysRevB.83.045201.

[32] W.A. Kamitakahara, C.M. Soukoulis, H.R. Shanks, U. Buchenau, G.S. Grest, Vibrational spectrum of amorphous silicon: Experiment and computer simulation, Phys. Rev. B 36 (1987) 6539–6542, http://dx.doi.org/10.1103/PhysRevB.36.6539.

[33] A. Gautam, Y.G. Lee, C. Ugwumadu, K. Nepal, S. Nakhmanson, D.A. Drabold, Site-projected thermal conductivity: Application to defects, interfaces, and homogeneously disordered materials, Phys. Status Solidi ( RRL) – Rapid Res. Lett. (ISSN: 1862-6270) (2024) http://dx.doi.org/10.1002/pssr.202400306.

[34] A.P. Thompson, H.M. Aktulga, R. Berger, D.S. Bolintineanu, W.M. Brown, P.S. Crozier, P.J. in 't Veld, A. Kohlmeyer, S.G. Moore, T.D. Nguyen, R. Shan, M.J. Stevens, J. Tranchida, C. Trott, S.J. Plimpton, LAMMPS - a flexible simulation tool for particle-based materials modeling at the atomic, meso, and continuum scales, Comput. Phys. Comm. 271 (2022) 108171, http://dx.doi.org/10.1016/j.cpc.2021.108171.

[35] M.H. Brodsky, M. Cardona, J.J. Cuomo, Infrared and Raman spectra of the silicon-hydrogen bonds in amorphous silicon prepared by glow discharge and sputtering, Phys. Rev. B 16 (1977) 3556–3571, http://dx.doi.org/10.1103/PhysRevB.16.3556.

[36] A.A. Langford, M.L. Fleet, B.P. Nelson, W.A. Lanford, N. Maley, Infrared absorption strength and hydrogen content of hydrogenated amorphous silicon, Phys. Rev. B 45 (1992) 13367–13377, http://dx.doi.org/10.1103/PhysRevB.45.13367.

[37] C.W. Rella, M. van der Voort, A.V. Akimov, A.F.G. van der Meer, J.I. Dijkhuis, Localization of the Si–H stretch vibration in amorphous silicon, Appl. Phys. Lett. 75 (19) (1999) 2945–2947, http://dx.doi.org/10.1063/1.125196.